\documentclass[conference]{IEEEtran}

\usepackage{hyperref}
\usepackage{graphicx}
\usepackage{xspace}
\usepackage{todonotes}
\usepackage{subcaption}

\usepackage{pifont}

\newcommand{\code}[1]{\textsf{\footnotesize#1}\xspace}

\newcommand{\squeeze}[1]{\vspace*{-#1mm}}
\renewcommand{\squeeze}[1]{}

\newcommand{\vulasreposapprox}{220\xspace}
\newcommand{\vulasbugsapprox}{540\xspace}

\newcommand{\vulascommitsapprox}{1100\xspace}

\newcommand{\inlinepar}[1]{\noindent\textbf{#1}}

\newcommand{\toprule}{\hline}
\newcommand{\midrule}{\hline}
\newcommand{\bottomrule}{\hline}

\newcommand{\as}[1]{{\noindent\textsf{\color{blue}AS:~{#1}}} \addcontentsline{tdo}{todo}{#1}}
\newcommand{\mb}[1]{{\noindent\textsf{\color{blue}MB:~{#1}}} \addcontentsline{tdo}{todo}{#1}}
\renewcommand{\as}[1]{}
\renewcommand{\mb}[1]{}

\title{\LARGE \bf A Practical Approach to the Automatic Classification of Security-Relevant Commits}
\author{Antonino Sabetta and Michele Bezzi \\
SAP Security Research \\
 {\tt\small \{antonino.sabetta,michele.bezzi\}@sap.com}
}

\begin{document}

\begin{figure*}
\begin{minipage}{\textwidth}
\begin{center}
{\LARGE \bf A Practical Approach to the \\[3mm] Automatic Classification of Security-Relevant Commits}\\[4mm]
{\Large [PRE-PRINT]} \\[8mm]
Antonino Sabetta and Michele Bezzi \\
SAP Security Research \\
{\tt\small \{antonino.sabetta,michele.bezzi\}@sap.com}
\end{center}

\vspace{10mm}
\noindent\textsc{Abstract}
\input{abstract}
\vspace{15mm}
\hrule
\vspace{10mm}
\begin{center}
{\Large Citing this paper}
\end{center}

This is a pre-print of the paper that appears in the proceedings of the
34th IEEE International Conference on Software Maintenance and Evolution
(ICSME) 2018.

If you wish to cite this work, please refer to it as follows:

\begin{verbatim}
@INPROCEEDINGS{sabetta2018icsme,
  author={Antonino Sabetta and Michele Bezzi},
  booktitle={34th IEEE International Conference
	           on Software Maintenance and Evolution (ICSME)},
  title={A Practical Approach to the Automatic
	       Classification of Security-Relevant Commits},
  year={2018},
  month={Sept},
}
\end{verbatim}
\vspace{5mm}
\hrule

\end{minipage}
\end{figure*}

\maketitle

\thispagestyle{empty}
\pagestyle{empty}

\begin{abstract}
The lack of reliable sources of detailed information on the
vulnerabilities of open-source software (OSS) components is a major obstacle to maintaining a secure software supply chain and an effective vulnerability management process. Standard sources of advisories and vulnerability data, such as the National Vulnerability Database (NVD), are known to suffer from poor coverage and inconsistent quality.

To reduce our dependency on these sources, 
we propose an approach that uses machine-learning to analyze source code repositories and to automatically identify commits that are \emph{security-relevant} (i.e., that are likely
to fix a vulnerability). We treat the source code changes introduced by commits as documents written in natural language, classifying them using standard document classification methods.

Combining independent classifiers  that use information from different facets of commits, our method can yield high precision (80\%) while ensuring acceptable recall (43\%).
In particular, the use of information extracted from the \emph{source code changes} yields a substantial improvement over the best known approach in state of the art, while requiring a significantly smaller amount of training data and employing a simpler architecture.

\end{abstract}

\section{Introduction}
\label{sec:intro}


The software industry is facing the difficult challenge of reconciling the
\emph{opportunity} represented by the availability of mature, high-quality open-source software (OSS)
components, with the \emph{problem} of maintaining a secure OSS supply chain and an effective vulnerability management process.
The lack of comprehensive, consistent, and up-to-date sources of vulnerability data represents a fundamental obstacle to tackle this challenge.

The current \emph{de-facto} standard source of vulnerability data, the \emph{National Vulnerability Database} (NVD)\footnote{\url{https://nvd.nist.gov/}}, is known to suffer from poor coverage: according to a recent study, as many as 25\% of the OSS projects \emph{silently fix} vulnerabilities without ever disclosing them in an official advisory~\cite{snyk_state_oss_sec_2017}. The same study reports that only 10\% of the OSS vulnerabilities receive a CVE (Common Vulnerabilities and Exposures) identifier and are published on the NVD.
\emph{Silent fixes} make it hard (if at all possible) for the clients
of the affected OSS component to assess the urgency of upgrading to a newly released version, because upgrades that only bring new features or functional improvements are made indistinguishable from those that (also) correct security flaws.

Furthermore, even the vulnerability advisories (CVEs), that \emph{will eventually} be posted in the NVD, follow a process that has a somewhat \emph{unpredictable timing}.
In practice, the interval between the
moment a vulnerability of an OSS component becomes \emph{known to the public} (e.g., because a commit fixing it
is pushed to the source code repository of the project) and the \emph{actual availability of a new release} that includes that fix, can last a few days or even weeks (sometimes months). During such time-frame, a malicious party that would examine the code commits in an open-source
project, could infer the existence of a security issue and exploit it before a fixed release is available through standard distribution channels.
This risk is pushing the industry towards monitoring source code repositories in real-time and automatically, to detect new security-relevant changes in a timely manner, without the need of waiting for official advisories to be published.



Manually linking a vulnerability and the commit(s) that
fix it requires considerable manual effort and expert knowledge, and is therefore expensive and error-prone. This problem motivated research
on automated methods that could alleviate the cost of this activity and of vulnerability analysis in general~\cite{Perl:2015:VFP:2810103.2813604,srcclr-esecfse-2017}.

In this paper we present a novel approach to the automated identification
of source code changes that are \emph{security-relevant}.
Our approach relies only on data available in code repositories, i.e., source code commits and their log messages, without the need of accessing vulnerability advisories. It allows a timely identification of security fixes, providing the necessary details to support a precise assessment of the impact of a OSS vulnerability using tools such as the one proposed by Ponta et al.~\cite{ponta2018}. To do so, inspired by the \emph{naturalness} hypothesis~\cite{allamanis2017survey,Hindle:2012:NS:2337223.2337322}, we
treat \emph{source code} changes as \emph{documents written in natural language} and we classify them using standard document classification methods.

We show that using information extracted directly from the source code, and combining independent classifiers suitably tuned, our method can yield high precision (80\%) while ensuring a relatively good recall (43\%). This result represents a substantial improvement over the best known approach in state of the art~\cite{srcclr-esecfse-2017}, while requiring a significantly smaller amount of training data and using a simpler architecture.


The remainder of the paper is organized as follows. Sec.~\ref{sec:relwork}
positions our contribution with respect to the existing literature. Sec.\ref{sec:approach} presents our approach to commit classification, whose quantitative evaluation is presented in Sec.~\ref{sec:evaluation}.
In Sec.~\ref{sec:lessons}  we elaborate on some
important aspects of commit classification in practice. Finally, Sec.~\ref{sec:conclusion} concludes the paper outlining the research directions
we intend to pursue in the future.

\section{Related Work}
\label{sec:relwork}

The research on applying machine learning to source code analysis has been very active in recent years.
Recent surveys of the field can be found in~\cite{Ghaffarian:2017:SVA:3135069.3092566} and in~\cite{allamanis2017survey}, where the emphasis is on approaches that, like ours does, rely on the \emph{naturalness of software}~\cite{Hindle:2012:NS:2337223.2337322}.

Concerning the concrete problem of \emph{commit classification}, the closest work to ours is~\cite{srcclr-esecfse-2017}, which presents an approach that uses an ensemble of six different models whose output is combined with an additional logistic regression model. Differently from us, the method of~\cite{srcclr-esecfse-2017} classifies a commit based only on its log message, whereas we also consider information extracted from the source code changes that it introduces.
Their architecture is more complex than ours, and uses more sophisticated word representation method (word embeddings, instead of plain \emph{bag of words}). The training in~\cite{srcclr-esecfse-2017} was done on a dataset of over 12400 commits (whereas we only used 2715). A comparison of their experimental results with ours is presented in Sec.~\ref{sec:evaluation}. The classification of bug tracking tickets, also addressed in~\cite{srcclr-esecfse-2017} is out of the scope of our work.

Vulnerability datasets like the one we used are a precious (and scarce) resource. Perl et al.~\cite{Perl:2015:VFP:2810103.2813604},
as part of their work on finding \emph{Vulnerability Contributing Commits} (VCC),
have produced what they claim to be the largest (public) database of mappings between CVEs affecting OSS and the
commits that \emph{introduced} the vulnerability. Their methodology, although automated, only considers CVEs that include an explicit reference to the corresponding commit(s); in our experience, these references are rather rare and often inaccurate. The dataset of~\cite{Perl:2015:VFP:2810103.2813604} includes 640 commits, corresponding to 718 CVEs, affecting 66 C/C++ projects.

Commercial vendors such as SourceClear (\url{http://sourceclear.com}) and Snyk (\url{http://snyk.io/vuln}) provide vulnerability data through their Web pages; Snyk also publishes monthly dumps of their databases on GitHub.
These resources could be used to replicate our study.

\section{Commit Classification}
\label{sec:approach}

Our approach to commit classification revolves around the following two key ideas.

First, we use conventional machine-learning algorithms and \emph{natural language processing} (NLP) methods. Differently from other works (such as~\cite{srcclr-esecfse-2017}) that concentrate on classifying commits based solely on the commit log message, we propose a novel way to treat the lines of code modified by a commit (in the following referred to as the \emph{patch}) as a \emph{text document} containing terms from the natural language. To do so, we extract the terms that the developer used to name entities in the source code (e.g., names of classes, functions, parameters, variables, constants etc.): the result is a document to which standard NLP methods can be applied.

Second, we are guided by the practical need, observed in our day-to-day work in the security department at \textsf{\small[COMPANY NAME]}, of ensuring high precision, and avoid wasting precious resources to deal with false-positives. To achieve this goal, we construct and train two different independent models, each considering a different aspect of the commit (the log message and the patch, respectively), tuning each of them for high-precision.  We experimented with different models (logistic regression, random forests, support vector machines, among others), and with different combinations of parameter settings (e.g., different kernels and regularization methods, and other model-specific parameters).
Due to space constraints, we omit here a discussion of the performance
obtained with each combination. We observed that SVM performs consistently
better than the other methods in our tests, and in particular, we found
that a linear kernel and L2 regularization provided both good predictive performance and acceptable training time.
We do not claim this is a general result, but we did observe that the impact of the choice of the model seems to have a lesser effect on predictive performance compared to the impact produced by different choices of feature sources and data pre-processing strategies.

When used separately, each of such models yields a relatively low recall. To compensate for this effect,
we combine the two models in a \emph{joint} model, implementing a simple \emph{voting mechanism} that flags a commit as security-relevant as soon as \emph{at least one of the two} models reports it as security-relevant. Because each model is trained on different facets of commits, their predictions are practically \emph{independent}, and therefore the likelihood that the joint model produces a false negative is reduced, thus increasing recall.

In the following we describe the two models (\code{msg} and \code{patch}) that are combined in the \code{joint} model.

\inlinepar{A)~Log Message Classifier (\code{msg}).} Using the well-known \emph{bag of words} approach, after some pre-filtering (stopword removal, normalization, removal of special characters, stemming) we transformed each log message into a vector of word counts and we trained a linear SVM model.

While this is a well-understood, basic document classification exercise, it comes with its own challenges. First, log messages can be rather short, and thus contain little information for the classifier to exploit. Second, by examining manually hundreds of security fixes, we observed that, depending on the community and on the specific project at hand, security fixes
can be quite difficult to distinguish (even for a human expert) based only on the commit message.

Ideally, one would expect that the log message should clearly indicate whether the commit at hand is meant to correct a security flaw. However, in practice, we observed that many projects avoid explicit log messages when pushing security fixes, possibly because that would make it easier for an attacker to infer what the vulnerability is, and to exploit it while a release embedding the fix is not ready yet.


\inlinepar{B)~Patch Classifier (\code{patch}).} Along the lines of~\cite{Hindle:2012:NS:2337223.2337322}, this approach stems
from the intuition that the choice of identifier names in source code
(that is the terms that the developer uses to name classes, methods,
variables) is an essential means to convey \emph{semantics} and thus
can convey information that can be used for classification.

Based on this intuition, we treat the source code changes themselves as a plain text document, and we reduce the problem of classifying code changes to the problem of classifying a document containing natural language terms. Once reduced to such a document, code changes can be represented as \emph{bags of words} and be treated with standard text-classification methods (similar to the log message classifier, except for a different preprocessing phase (see next section).

\section{Evaluation}
\label{sec:evaluation}




\subsection{Dataset}
\label{sec:data-prep}

For our studies we used a labeled dataset of known open-source vulnerabilities
constructed while operating an open-source vulnerability scanner,
described in~\cite{ponta2018}, developed and adopted at SAP for internal use.
Differently from~\cite{Perl:2015:VFP:2810103.2813604}, who collected commits
that \emph{introduce} vulnerabilities, our dataset
contains commits that implement \emph{fixes} to vulnerabilities.
Its data covers \vulasreposapprox open-source Java projects that are
used in SAP software (either products or internal tools) and contains
around \vulascommitsapprox security-relevant commits
corresponding to the fixes for \vulasbugsapprox vulnerabilities (e.g., CVEs).
For our study we used a subset of 660 such commits from 152 repositories.

The data preparation
phase includes several ad-hoc steps of little technical interest, but whose
impact on the classification results is significant: we summarize here the main ones in the hope of encouraging studies that replicate (and possibly improve) our results. After skipping invalid commits, we removed duplicates (commits with different identifier but identical content), obtaining 456 unique commits, assigned to the \emph{positive} (pos) class.

We then augmented the dataset with instances of the ``negative'' class (non-security-relevant): for each positive instance $p$ from repository $R$, we took $k$ random commits $n_1,\ldots,n_k$ from $R$ and, under the assumption that security-relevant commits are rare compared to other types of commits, we treated these as ``negative'' examples. To check the validity of this assumption, and to avoid including obvious outliers (extremely large, empty, or otherwise invalid commits), we performed a manual review of these commits, with the help of ad-hoc scripts and pattern matching (similarly to~\cite{srcclr-esecfse-2017}).
We chose $k=5$ to get approximately $(1 + 5) \times 456$ commits overall.
After the manual review, we obtained a dataset of 2715 instances (456 of the positive class, and 2259 of the negative class).

This raw dataset was further processed to remove stop-words, filter non-alphanumeric characters and normalize the text data. In particular, for the patch data, we removed context lines, keeping the actual changed lines, and splitting composite tokens (e.g., camel-case identifiers).


\begin{figure}[tb]
	\begin{center}
	\includegraphics[width=0.8\linewidth]{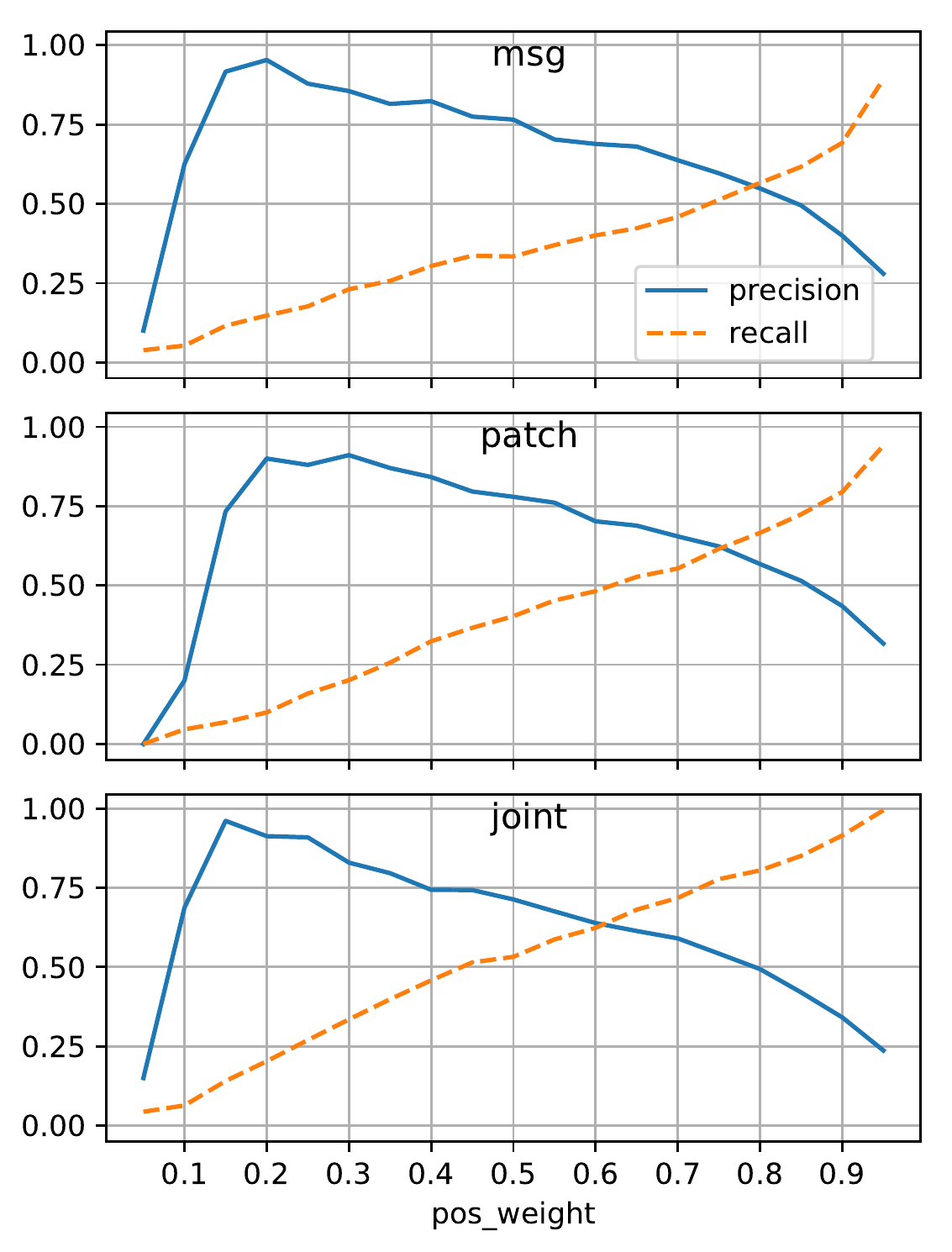}
    \vspace{-3mm}
  	\caption{Evaluation results\\[-6mm]}
	\label{fig:evaluation}
    \end{center}
\end{figure}

\subsection{Results}
We used 10-fold cross-validation to evaluate the predictive performance of our models in terms of precision and recall, and we compared them to the best results in the state of the art. The outcome of our experiments is summarized in Fig.~\ref{fig:evaluation}, for the two models \emph{log message classifier} (\code{msg}) and \emph{patch classifier} (\code{patch}) considered independently, as well as for their combination (\code{joint}). The x-axis represents the weight $w_p$ assigned to the misclassification penalty for the positive class (the negative class gets $w_n = 1-w_p$).

These plots illustrate clearly the trade-off between precision and recall. Depending on the \emph{operating point} (controlled by adjusting the weights $w_p, w_n$), these models  can be tuned to obtain the desired level of precision, at the expense of lower recall, or vice-versa.

The two left-most columns of the table below contain the different performance figures reported by~\cite{srcclr-esecfse-2017}, when comparing their approach with prior work or when describing the model they deployed in production.
For comparison, in the three right-most columns we report the precision obtained with each of our models (\code{msg}, \code{patch}, \code{joint}) for each level of recall fixed in~\cite{srcclr-esecfse-2017}.

\begin{center}
\footnotesize
\begin{tabular}{c|c|ccc}
\toprule
 rec.  &  prec.~\cite{srcclr-esecfse-2017} & prec. (\code{msg}) &  prec. (\code{patch})	& prec. (\code{joint}) \\
\midrule
0.50	&	0.51	&	0.58	&	0.71	&	0.74	\\
0.72	&	0.34	&	0.42	& 	0.51	&	0.57	\\
0.76	&	0.34	&	0.39	&	0.45	&	0.56	\\
\bottomrule
\end{tabular}
\end{center}

As we explain in the next section, practical considerations stemming from industrial practice
make us emphasize precision rather than recall.
The table below shows the recall figures obtained when fixing precision at different levels equal to or higher than $0.80$.

\begin{center}
\footnotesize
\begin{tabular}{c|ccc}
\toprule
prec. & rec. (\code{msg}) &  rec. (\code{patch})	& rec. (\code{joint})  \\
\midrule
0.95	&	0.15	&	0.14	& 	0.20  \\
0.90	&	0.19	&	0.18	&	0.25  \\
0.85	&	0.21	&	0.23	&	0.31  \\
0.80	&	0.30	&	0.29	&	0.43  \\
\bottomrule
\end{tabular}
\end{center}

These figures show that, when tuning our models for high-precision, we still obtain useful levels of recall, which makes our approach suitable for real-world industrial use.

\section{Commit Classification in Practice}
\label{sec:lessons}

We briefly discuss here two important aspects of practical commit classification: the choice of reference performance metrics, and the generalization of our approach to projects written in languages other than Java.

\inlinepar{The Importance of Precision.} In an industrial scenario,
the security-relevant commits
identified through automated classification need to go through the manual assessment of a human expert, who decides whether they should be added to the corporate database of vulnerabilities that is used by automated vulnerability scanners to check hundreds of projects across the company. In this context, any approach that would overwhelm the expert with too many false positives would be unsustainable; in our experience, precision figures below $0.80$ are impractical,
because the number of commits analyzed daily can quickly become unwieldy.

\inlinepar{Cross-Language Application.} While our approach is inherently (programming-)language independent, its predictive performance could change when our model is applied to projects written in programming languages other than the one covered by the training data.

In an initial attempt to study this aspect, we analyzed the 30 most recent vulnerabilities affecting \emph{Ruby} projects published on \url{snyk.io} at the time of writing.
For five of the 30 vulnerabilities, we could not identify a corresponding commit. On the other hand, some vulnerabilities are linked to multiple commits, and some commits contain the same changes. As a result, we obtained a dataset of 23 unique commits of the ``positive'' class, from 15 distinct repositories. Using the same technique as in Sec.~\ref{sec:data-prep}, we derived a new validation dataset (\code{Snyk-Ruby} in the following) containing 224 additional ``negative'' samples (summing up to 247 commits).

We classified these commits with the joint predictor model to  that we had previously trained on Java-only data, which resulted in a noticeable improvement of performance compared to the result obtained in the cross-validation using the SAP dataset. For comparison, on the \code{Snyk-Ruby} dataset we obtained a precision of 82\% with a recall of 61\%, whereas the best result we obtained on the SAP dataset (Java) was 80\% precision and 43\% recall.
A similar result was reported in~\cite{srcclr-esecfse-2017}, with a increase in performance
when using a model trained on Java to classify projects coded in Ruby or Go.
%
%

Based on our experience and some anecdotal evidence, we believe that these differences in performance might depend on  the practices followed in different communities and even in different projects: the information included  in log messages and the style of fixes (e.g., the amount and content of comments included with security fixes) could make it substantially easier (or harder) to classify them correctly.


While, of course, \emph{no generality} can be claimed for the results of this simple exercise, they suggest that the effect of community-specific (and project-specific) practices on classification performance cannot be neglected.
A systematic study of this problem is in our plans for future research.

\section{Conclusion}
\label{sec:conclusion}

Motivated by the need to support open-source vulnerability management in an enterprise environment, we tackled the problem of identifying security-relevant commits in source code repositories using machine learning.

By translating the problem of classifying \emph{code changes} to the problem of classifying a \emph{document containing natural language terms}, we could apply classification models that require no sophisticated code analysis and are agnostic of the programming language. By combining relatively simple models optimized for high precision, we obtained classification capabilities that are suitable for industrial use. Our approach outperforms state-of-the-art approaches~\cite{srcclr-esecfse-2017}, while employing a computationally less expensive model and requiring a considerably smaller training dataset.

We do believe that additional improvements are possible, and that further research in model selection and optimization is necessary. We expect more \emph{modern} approaches (e.g., word embeddings, recurrent neural networks) to improve performance even further and we plan to investigate their applicability. A key conclusion of this work is that, as a consequence of the \emph{naturalness} of source code~\cite{allamanis2017survey}, even using basic NLP methods and without recurring to sophisticated code-analysis techniques, effective and relatively inexpensive commit classification can be achieved. A promising direction we are exploring at the time of writing is the classification of a given commit together with other resources (bug tracking tickets, advisories, mailing list discussions) that \emph{link} to it~\cite{the-missing-links-bugs-and-bug-fix-commits}. We will report on our progress in this direction in a future paper.




\subsection*{Acknowledgements.}  We would like to acknowledge the insightful comments of the anonymous reviewers who contributed significantly to improve this paper.
We are also grateful to Ivan Pashchenko, Sule Kharaman, and Jamarber Bakalli for their help with the validation of our approach, and to our colleagues Henrik Plate and Serena E. Ponta for their comments on early drafts of this work.

%
%


\bibliographystyle{IEEEtran}
\bibliography{ml}

\begin{thebibliography}{1}
\providecommand{\url}[1]{#1}
\csname url@samestyle\endcsname
\providecommand{\newblock}{\relax}
\providecommand{\bibinfo}[2]{#2}
\providecommand{\BIBentrySTDinterwordspacing}{\spaceskip=0pt\relax}
\providecommand{\BIBentryALTinterwordstretchfactor}{4}
\providecommand{\BIBentryALTinterwordspacing}{\spaceskip=\fontdimen2\font plus
\BIBentryALTinterwordstretchfactor\fontdimen3\font minus
  \fontdimen4\font\relax}
\providecommand{\BIBforeignlanguage}[2]{{%
\expandafter\ifx\csname l@#1\endcsname\relax
\typeout{** WARNING: IEEEtran.bst: No hyphenation pattern has been}%
\typeout{** loaded for the language `#1'. Using the pattern for}%
\typeout{** the default language instead.}%
\else
\language=\csname l@#1\endcsname
\fi
#2}}
\providecommand{\BIBdecl}{\relax}
\BIBdecl

\bibitem{snyk_state_oss_sec_2017}
\BIBentryALTinterwordspacing
Snyk.io, ``The state of open-source security,'' 2017. [Online]. Available:
  \url{https://snyk.io/stateofossecurity/pdf/The%20State%20of%20Open%20Source.pdf}
\BIBentrySTDinterwordspacing

\bibitem{Perl:2015:VFP:2810103.2813604}
\BIBentryALTinterwordspacing
H.~Perl, S.~Dechand, M.~Smith, D.~Arp, F.~Yamaguchi, K.~Rieck, S.~Fahl, and
  Y.~Acar, ``Vccfinder: Finding potential vulnerabilities in open-source
  projects to assist code audits,'' in \emph{Proceedings of the 22Nd ACM SIGSAC
  Conference on Computer and Communications Security}, ser. CCS '15.\hskip 1em
  plus 0.5em minus 0.4em\relax New York, NY, USA: ACM, 2015, pp. 426--437.
  [Online]. Available: \url{http://doi.acm.org/10.1145/2810103.2813604}
\BIBentrySTDinterwordspacing

\bibitem{srcclr-esecfse-2017}
\BIBentryALTinterwordspacing
Y.~Zhou and A.~Sharma, ``Automated identification of security issues from
  commit messages and bug reports,'' in \emph{Proceedings of the 2017 11th
  Joint Meeting on Foundations of Software Engineering}, ser. ESEC/FSE
  2017.\hskip 1em plus 0.5em minus 0.4em\relax New York, NY, USA: ACM, 2017,
  pp. 914--919. [Online]. Available:
  \url{http://doi.acm.org/10.1145/3106237.3117771}
\BIBentrySTDinterwordspacing

\bibitem{ponta2018}
S.~E. Ponta, H.~Plate, and A.~Sabetta, ``Beyond metadata: Code-centric and
  usage-based analysis of known vulnerabilities in open-source software,'' in
  \emph{2018 IEEE International Conference on Software Maintenance and
  Evolution (ICSME)}, Sept 2018.

\bibitem{allamanis2017survey}
M.~Allamanis, E.~T. Barr, P.~Devanbu, and C.~Sutton, ``A survey of machine
  learning for big code and naturalness,'' \emph{arXiv preprint
  arXiv:1709.06182}, 2017.

\bibitem{Hindle:2012:NS:2337223.2337322}
\BIBentryALTinterwordspacing
A.~Hindle, E.~T. Barr, Z.~Su, M.~Gabel, and P.~Devanbu, ``On the naturalness of
  software,'' in \emph{Proceedings of the 34th International Conference on
  Software Engineering}, ser. ICSE '12.\hskip 1em plus 0.5em minus 0.4em\relax
  Piscataway, NJ, USA: IEEE Press, 2012, pp. 837--847. [Online]. Available:
  \url{http://dl.acm.org/citation.cfm?id=2337223.2337322}
\BIBentrySTDinterwordspacing

\bibitem{Ghaffarian:2017:SVA:3135069.3092566}
\BIBentryALTinterwordspacing
S.~M. Ghaffarian and H.~R. Shahriari, ``Software vulnerability analysis and
  discovery using machine-learning and data-mining techniques: A survey,''
  \emph{ACM Comput. Surv.}, vol.~50, no.~4, pp. 56:1--56:36, Aug. 2017.
  [Online]. Available: \url{http://doi.acm.org/10.1145/3092566}
\BIBentrySTDinterwordspacing

\bibitem{the-missing-links-bugs-and-bug-fix-commits}
\BIBentryALTinterwordspacing
A.~Bachmann, C.~Bird, F.~Rahman, P.~Devanbu, and A.~Bernstein, ``The missing
  links: Bugs and bug-fix commits,'' in \emph{SIGSOFT '10/FSE-18: Proceedings
  of the 16th ACM SIGSOFT Symposium on Foundations of Software
  Engineering}.\hskip 1em plus 0.5em minus 0.4em\relax Association for
  Computing Machinery, Inc., November 2010. [Online]. Available:
  \url{https://www.microsoft.com/en-us/research/publication/the-missing-links-bugs-and-bug-fix-commits/}
\BIBentrySTDinterwordspacing

\end{thebibliography}

\end{document}